\documentclass[11pt, onecolumn, draftcls, singlespace ]{IEEEtran}
\usepackage{cite}
\usepackage{algorithm2e}
\usepackage{amssymb}
\usepackage{enumerate}
\usepackage{graphicx}
\usepackage{amssymb}
\usepackage{amsbsy}
\usepackage{float}
\usepackage{balance}
\usepackage{url}
\usepackage{bm}
\usepackage{array}
\usepackage{varioref}
\usepackage[normalem]{ulem}
\usepackage{color}
\usepackage{empheq}
\usepackage{flushend}
\begin{document}

\title{Weighted {Fair}  Multicast Multigroup Beamforming under Per-antenna Power Constraints}
\author{Dimitrios Christopoulos$^\star$, \IEEEmembership{Student Member, IEEE,} Symeon Chatzinotas, \IEEEmembership{Member, IEEE,}\\ and Bj\"orn Ottersten, \IEEEmembership{Fellow, IEEE}
\thanks{The authors are with the SnT--University of Luxembourg. \ \ \ \ \ \ \ \ \ \ \ \ \ \ \ \ \ \ \ \ \ \ \ \ Email: \textbraceleft dimitrios.christopoulos, symeon.chatzinotas, bjorn.ottersten\textbraceright@uni.lu}
  \thanks{This work was   partially supported by the National Research Fund, Luxembourg under the projects  ``$\mathrm{ CO^{2}SAT}$''  and ``$\mathrm{ SemiGod}$''. {Part of this work has been accepted for presentation in the IEEE ICC 2014, Sydney AUS.} }
%

%
}
\maketitle


\begin{abstract}
A multi-antenna transmitter that conveys independent sets of common data to distinct groups of  users is  considered.   This model is known as physical layer multicasting to multiple co-channel groups. In this context, the practical constraint of a maximum permitted  power level  radiated by each antenna is addressed.   The  per-antenna power constrained system is optimized in a  maximum fairness sense with respect to predetermined quality of service weights. In other words,   the worst scaled user is boosted  by  maximizing  its weighted  signal-to-interference plus noise ratio. A detailed solution to tackle  the weighted max-min fair multigroup multicast problem under per-antenna power constraints is therefore derived.   The implications of the novel constraints   are investigated via  prominent applications and paradigms. What is more, robust per-antenna constrained multigroup multicast beamforming solutions are proposed. Finally, an extensive performance evaluation quantifies the gains of the proposed algorithm over existing solutions and exhibits its accuracy over per-antenna power constrained systems.
\end{abstract}
 \section{{Introduction \& Related Work}}
  The spatial degrees of freedom offered by multiple antenna arrays are a valuable interference mitigation resource.  Advanced signal processing techniques are currently  employed to boost the performance of the multi-antenna transmitters without compromising the complexity of single antenna receivers.
These   beamforming (or equivalently precoding) techniques efficiently manage the co-channel interferences to achieve the targeted service requirements  (Quality of Service--$\mathrm{QoS}$ targets). As a result,   the available spectrum can be aggressively reused towards increasing the system throughput.

The optimal downlink transmission strategy in the sense of minimizing the total transmit power whilst guaranteing specific $\mathrm{QoS}$ targets at each user,    was given in \cite{Bengtsson2001,bengtsson1999}.
{Therein, the  tool of Semi-Definite Relaxation ($\mathrm{SDR}$) reduced the non-convex quadratically constrained quadratic  problem ($\mathrm{QCQP}$)  into a relaxed semi-definite programming  instance by changing the optimization variables and disregarding the unit-rank constraints over the new variable. The solution of the relaxed problem was proven to be optimal.  {The multiuser downlink beamforming problem in terms of maximizing the minimum $\mathrm{SINR}$,  was optimally solved in \cite{Schubert2004}. The goal of the later  formulation is to increase the fairness of the system by boosting the  $\mathrm{SINR}$ of the  user that is further away from a targeted performance. Hence,  the problem is commonly referred to as \textit{max--min fair}.  In \cite{Schubert2004}, this problem was solved using the principles of uplink/downlink duality. Therein,  \textit{Schubert and Boche} developed a strongly convergent iterative alternating optimization algorithm for the equivalent uplink problem.   In the same work, the power minimization problem of \cite{Bengtsson2001} was also solved  by acknowledging its inherent connection with the max-min fair problem. Consequently, a significantly less complex  framework to solve the optimal beamforming problem was established. }Extending these works, the practical per-antenna  power constraints $\mathrm{  (PAC)}$ were considered  in \cite{Yu2007}. {Generalized power constraints, including sum power, per-antenna power and per-antenna array power constraints were considered in \cite{Dartmann2013}, where the proposed max-min fair solution was  derived on an  extended duality framework. This framework accounted for both instantaneous and long term channel state information ($\mathrm{CSI}$).} $\mathrm{PAC}$s are motivated from the practical implementation of systems that rely on precoding. The lack of flexibility in sharing energy resources amongst the antennas of the transmitter is usually the case, since a  common practice in multi-antenna systems is the use of individual amplifiers per  antenna. Despite the fact that flexible amplifiers could be incorporated in multi-antenna transmitters, specific communication systems cannot afford this design. Typical per antenna power limited systems can be found in  multibeam satellite communications \cite{Christopoulos2013AIAA}, where flexible on board payloads are difficult to implement and in cooperative  multicell systems (also known  as distributed antenna systems, $\mathrm{DAS}$), where the  physical co-location of the transmitting elements is not a requisite and hence power sharing might be infeasible.

A fundamental  consideration of the aforementioned works is that  independent  data is addressed to multiple users. However, the new generation of multi-antenna communication standards has to adapt  the physical layer design to the needs of the higher network  layers.  Examples of  such cases include  highly demanding applications (e.g. video broadcasting) that stretch the throughput limits  of multiuser broadband systems.
In this direction, physical layer ($\mathrm{PHY}$) multicasting
has the potential to efficiently address the nature of future traffic demand and  has become part of the new generation of communication standards. $\mathrm{PHY}$ multicasting is also relevant for the application of beamforming  without changing the framing structure of standards.{Such a scenario can be found in satellite communications where the communication standards are  optimized to cope with  long propagation delays and guarantee scheduling efficiency by framing multiple users per transmission \cite{Christopoulos2013AIAA,Christopoulos2014_ASMS}.}

In \cite{Sidiropoulos2006},   the NP-hard multicast problem was accurately approximated by $\mathrm{SDR}$ and Gaussian randomization.
The natural  extension of the multicast concept lies in assuming multiple interfering  groups of users.
A unified framework for physical layer multicasting to multiple co-channel groups, where independent sets of common data are transmitted to  groups of users by the multiple antennas, was given in \cite{Karipidis2005CAMSAP,Karipidis2008}. Therein, the  $\mathrm{QoS}$ and the fairness problems were formulated, proven NP-hard and solved for  the sum power constrained multicast multigroup case.  In parallel to \cite{Karipidis2005CAMSAP}, the independent work   of \cite{Gao2005}  involved complex dirty paper coding methods. Also, a convex approximation method was proposed in \cite{Pesavento2012b} that exhibits superior performance as the number of users per group grows. Finally, in \cite{Silva2009}
the multicast multigroup problem under $\mathrm{SPC}$, was solved based on approximations and uplink-downlink duality \cite{Schubert2004}.
{In the context of coordinated multicast multicell systems{\footnote{{ Coordinated multicell networks consist of connected base stations ($\mathrm{BS}$), with each $\mathrm{BS}$ serving a single multicast group, a case tackled in \cite{Xiang2013}. Extending this,  the methods presented herein can be applied in cooperative multicell  systems where all $\mathrm{BS}$s will jointly transmit to several multicast groups\cite{Chatzinotas_JWCOM}.}}},  max--min fair beamforming with per base-station ($\mathrm{BS}$) constraints has been considered in \cite{Xiang2013} where  each $\mathrm{BS}$ transmits to a single multicast group.   Hence, a power constraint over each precoder was imposed while  no optimization weights were considered.This formulation still considers power sharing amongst the multiple antennas at each transmitter.}

Towards deriving the optimal multigroup multicast precoders when a maximum limit is imposed on the  transmitted power of each antenna, a new optimization problem with  one constraint per transmit antenna needs to be formulated.
{Amid the extensive literature on multigroup multicast beamforming, the $\mathrm{PAC}$s have only been considered in \cite{Christopoulos2014_ICC}, where an   equally fair multicast multigroup solution is presented.} Extending these considerations, the present work accounts optimization weights.  Therefore,  a consolidated solution for the weighted max--min fair multigroup multicast beamforming under $\mathrm{PAC}$s is hereafter presented. The contributions of the present work  are summarized as follows
{\begin{itemize}
\item The $\mathrm{PAC}$ weighted fair multigroup multicast beamforming problem is formulated and solved.
\item Practical system design insights are given by examining the implications of the $\mathrm{PAC}$s on  multigroup multicast distributed antenna systems ($\mathrm{DAS}$), modulation constrained systems and  uniform linear array  ($\mathrm {ULA}$) transmitters.
\item A robust to erroneous $\mathrm{CSI}$ multigroup multicast design under  $\mathrm{PAC}$s is proposed.
\item The {performance  of the solution} is evaluated through extensive numerical results under various system setups.\end{itemize}
}

The rest of the paper is structured as follows.   The multigroup multicast system model is presented in Sec. \ref{sec: System Model} while the weighted  fair problem is formulated and solved in Sec. \ref{sec: problem}. In Sec. \ref{sec: performance}, the performance of the design is evaluated for various system setups along with a robust extension of the derived algorithm and a weighted multigroup multicast application paradigm. Finally,  Sec. \ref{sec: conclusions} concludes the paper.

{\textit{Notation}: In the remainder of this paper, bold face lower case and upper case characters denote column vectors  and matrices, respectively. The operators \(\left(\cdot\right)^\text{T}\), \(\left(\cdot\right)^\dag\), $|\cdot|$, ${\mathrm{Tr}\left(\cdot\right)}$ and \(||\cdot||_2, \) correspond to   the transpose, the conjugate transpose,  the absolute value, the {trace} and the Frobenius norm operations,  while $[\cdot]_{ij}  $  denotes the $i, j$-th element of a matrix.  The principal eigenvalue of a matrix $\mathbf X$ are denoted as $\lambda_{max}(\mathbf X)$. Calligraphic indexed characters denote sets}.

\section{System Model }\label{sec: System Model}
 Herein, the focus is on a multi-user ($\mathrm{MU}$) multiple input single output ($\mathrm{MISO}$) multicast system. Assuming a single transmitter, let   $N_t$ denote the number of transmitting elements  and  $N_{u}$ the  total number of users served.  The input-output analytical expression  will read as $y_{i}= \mathbf h^{\dag}_{i}\mathbf x+n_{i},$
where \(\mathbf h^{\dag}_{i}\) is a \(1 \times N_{t}\) vector composed of the channel coefficients (i.e. channel gains and phases) between the \(i\)-th user and the  \(N_{t}\) antennas of the transmitter, \(\mathbf x\) is the \(N_{t} \times 1\)  vector of the transmitted symbols and  \(n_{i}\) is the independent complex circular symmetric (c.c.s.) independent identically distributed (i.i.d) zero mean  Additive White Gaussian Noise ($\mathrm{AWGN}$)  measured at the \(i\)-th user's receive antenna.
Focusing  in a multigroup multicasting scenario,  let there be a total of $1\leq G \leq N_{u}$ multicast groups with  $\mathcal{I} = \{\mathcal{G}_1, \mathcal{G}_2, \dots  \mathcal{G}_G\}$ the collection of   index sets and $\mathcal{G}_k$ the set of users that belong to the $k$-th multicast group, $k \in \{1\dots G \}$. Each user belongs to only one group, thus $\mathcal{G}_i\cap\mathcal{G}_j=$\O ,$  \forall i,j \in \{1\cdots G\}$. Let $\mathbf w_k \in \mathbb{C}^{N_t \times 1}$ denote the precoding weight vector applied to the transmit antennas to beamform towards the $k$-th group.
The assumption of independent data transmitted to different groups renders the symbol streams $\{s_k\}_{k=1}^G$ mutually uncorrelated and the total power radiated from the antenna array is
\begin{align}
P_{tot} = \sum_{k=1}^ G \mathbf w_k{^\dag} \mathbf w_k 
\end{align}
The power radiated by each
antenna element is  a  linear combination of all precoders
\cite{Yu2007}:\begin{align}\label{eq: PAC}
P_n = \left[\sum_{k=1}^G \mathbf w_k \mathbf w_k^\dag \right]_{nn} 
\end{align}
where $n \in \{1\dots  N_t\}$ is the antenna index.
The fundamental difference between the $\mathrm{SPC}$  of \cite{Karipidis2008} and the proposed $\mathrm{PAC}$ is clear in  \eqref{eq: PAC}, where instead of one,  $N_t$ constraints are realized, each one involving all the precoding vectors.  A more general constraint formulation to model power flexibility amongst groups of antennas can be found in \cite{Zheng2011a}.
\section{Multicast Multigroup Beamforming with Per Antenna Power Constraints}\label{sec: problem}
\subsection{Weighted Max-Min Fair Formulation}
The $\mathrm{PAC}$ weighted max-min fair  problem   is defined as
\begin{empheq}[box=\fbox]{align}
\mathcal{F:}\    \max_{\  t, \ \{\mathbf w_k \}_{k=1}^{G}}  &t& \notag\\
\mbox{subject to } & \frac{1}{\gamma_i}\frac{|\mathbf w_k^\dag \mathbf h_i|^2}{\sum_{l\neq k }^G |\mathbf w_l^\dag\mathbf h_i|^2+\sigma_i^2 }\geq t, &\label{const: F SINR}\\
&\forall i \in\mathcal{G}_k, k,l\in\{1\dots G\},\notag\\
 \text{and to }\ \ \ \ & \left[\sum_{k=1}^G  \mathbf w_k\mathbf w_k^\dag  \right]_{nn}  \leq P_n, \label{eq: max-min fair power const1 }\\
 &\forall n\in \{1\dots N_{t}\},\notag
 \end{empheq}
 where $\mathbf w_k\in \mathbb{C}^{N_t}$ and $t \in \mathbb{R}^{+}$. Different service levels between the users can be acknowledged in this weighted  formulation.   Problem $ \mathcal{F}$ receives as inputs the  $\mathrm{PAC}$s vector $\mathbf p = [P_1, P_2\dots P_{N_t}]$ and the target $\mathrm{SINR}$s vector $\mathbf g = [\gamma_1,\gamma_2, \dots \gamma_{N_u}]$. {Its goal is to maximize the slack variable $t$ while keeping all $\mathrm{SINR}$s above this value. Thus, it constitutes a max-min problem that guarantees fairness amongst users.}   Following the common  in the literature notation for ease of reference, the  optimal objective value of $ \mathcal{F}$ is denoted as $t^*=\mathcal{F}(\mathbf g,  \mathbf p)$ and the associated optimal point as $\{\mathbf w_k^\mathcal{F}\ \}_{k=1}^{G}$. {Of particular interest is the case where the co-group users  share the same target i.e. $\gamma_i = \gamma_{k},\ \forall i \in\mathcal{G}_k, k\in\{1\dots G\}$.}

{\textit{Remark 1}: {The difference of the present formulation with respect to the weighted max-min fair problem with $\mathrm{SPC}$ presented in   \cite{Sidiropoulos2006,Karipidis2008} lies in  the $N_t$ power constraints over each individual radiating element. Additionally, this formulation differs from the coordinated multicell multicasting Max-Min formulation of \cite{Xiang2013} since {the constraint  is imposed on  the $n$-th   diagonal element of the summation of the correlation matrices of all precoders, while weights on each users } $\mathrm{SINR}$ are also inserted. On the contrary, in \cite{Xiang2013}, the imposed per base station constraints are translated to one power constraint per each precoder.
In the present work, weights to differentiate the $\mathrm{QoS}$ targets between users are also proposed.}}

\subsection{Per-antenna power  minimization}
{ The relation between the fairness  and the power minimization problems for the multicast multigroup case was firstly established in \cite{Karipidis2008}. As a result, by bisecting  the solution of the  $\mathrm{QoS}$ optimization, a solution to the weighted fairness problem can be derived. Nevertheless, fundamental differences between the existing formulations and problem $\mathcal{F}$ complicate the solution. In more detail, the per-antenna constraints are not necessarily met with equality (a discussion on this is also given in Sec. \ref{sec: power consumption}). Therefore, the fairness problem is no longer equivalent to the sum power minimization under  $\mathrm{QoS}$ constraints problem. Since the absence of a related, solvable  problem prohibits the immediate application of bisection,  a novel equivalent per-antenna power  minimization problem is proposed as}
\begin{empheq}[box=\fbox]{align}
\mathcal{Q:} \min_{\ r, \ \{\mathbf w_k \}_{k=1}^{G}}  &r& \notag\\
\mbox{subject to } & \frac{|\mathbf w_k^\dag \mathbf h_i|^2}{\sum_{l\neq k }^G | \mathbf w_l^\dag\mathbf h_i|^2 + \sigma^2_i}\geq \gamma_i, \label{const: Q SINR}\\
&\forall i \in\mathcal{G}_k, k,l\in\{1\dots G\},\notag\\
\text{and to} \ \ \ \ \ & \frac{1}{P_n} \left[\sum_{k=1}^G  \mathbf w_k\mathbf w_k^\dag \right]_{nn} \leq  r,\label{eq: PAC Q}\\
& \forall n\in \{1\dots N_{t}\}, \notag
 \end{empheq}
 with $r\in\mathbb{R^+}$. Problem $\mathcal{Q }$ receives as input  $\mathrm{SINR}$ constraints for  all users, defined   before as $\mathbf g $,  as well as the per antenna power constraint vector $\mathbf p$ of \eqref{eq: max-min fair power const1 }. {The introduction of the slack-variable  $r$,  a common practice in convex optimization \cite{convex_book}, constraints the power consumption of each and every antenna.   Subsequently, at the optimum $r^*$,   the maximum  power consumption out of all antennas is minimized} and this solution is denoted as $r^*=\mathcal{Q}(\mathbf g, \mathbf p)$. {The generic difference of the present min-max formulation and the formulation proposed in \cite{Xiang2013} lies in the per antenna constraint \eqref{eq: PAC Q}. Instead of constraining the  power of each antenna, the authors of \cite{Xiang2013} impose a constraint over each precoder that serves a common multicast group. In the case tackled herein, the number of constraints is increased from one to $N_t$, while  each constraint is a function of all multigroup precoders as the summation in \eqref{eq: PAC Q} reveals.}     The following claim reveals the relation between the described problems.

 \textit{Claim 1}: Problems $\mathcal{F}\ $ and  $\mathcal{Q}\ $ are related as follows
\begin{align}
1 = \mathcal{Q}\left(\mathcal{F}\left( \mathbf g, \mathbf p\right)\cdot\mathbf g, \mathbf p \right)\label{eq: equivalence 1}\\
t = \mathcal{F}\left(\mathbf g, \mathcal{Q}\left( t\cdot \mathbf g,\mathbf p\right)\cdot\mathbf p \right)\label{eq: equivalence 2}
 \end{align}
 \textit{Proof: }{Similar to the line of reasoning in \cite{Xiang2013} the above claims will be proven by contradiction.} Starting with   \eqref{eq: equivalence 1},  let $t^*=\mathcal{F}(\mathbf g,\ \mathbf p)$ denote the optimal value of $\mathcal{F}$ with associated variable $\{\mathbf w_k^F \}_{k=1  }^{G}$.
Also, let $\hat r=\mathcal{Q}\left( t^*\cdot\mathbf g,\ \mathbf p\right)$ be the optimal value of $\mathcal{Q}$ at the point  $ \{\mathbf w_k^Q \}_{k=1  }^{G}$. Then, assuming that $\hat r > 1$, the vectors $ \{\mathbf w_k^F \}_{k=1  }^{G}$ satisfy the feasibility criteria of  $\mathcal{Q}$ and produce a lower optimal value  thus contradicting the optimality of  $ \{\mathbf w_k^Q \}_{k=1  }^{G}$ and opposing the hypothesis. Alternatively, assuming that $\hat r < 1$ then the solutions $ \{\mathbf w_k^Q \}_{k=1  }^{G}$  can be scaled by the non-negative $ \hat r $. The vectors $ \{ \hat r \cdot\mathbf{ w}_k^Q \}_{k=1  }^{G}$ are feasible solutions to $\mathcal{F }$  which provide the same optimal objective value with however some remaining power budget. Therefore, the power could be scaled up until at least one of the $\mathrm{PAC}$s is satisfied with equality and a higher objective value would be derived thus again contradicting the hypothesis. Consequently,  $\hat r = 1$.
The same line of reasoning is followed to prove \eqref{eq: equivalence 2}. Let $r^*=\mathcal{Q}(t\cdot\mathbf g,\ \mathbf p)$ denote the optimal value of $\mathcal{Q}$ with associated solution $ \{\mathbf w_k^Q \}_{k=1  }^{G}$. Assuming that the optimal value of $\mathcal{F}$ under constraints scaled by the solution of $\mathcal{Q}$ is different, i.e. $\hat t= \mathcal{F}\left(\mathbf g , \ \mathcal{Q}\left(t\cdot \mathbf g ,\ \mathbf p\right)\cdot\mathbf p\right)$ with $ \{\mathbf w_k^F \}_{k=1}^{G}$, the following contradictions arise. In the case where $\hat t < t $, then the precoders   $ \{\mathbf w_k^Q \}_{k=1 }^{G}$  are feasible solutions to  $\mathcal{F}$ which lead to a higher minimum $\mathrm{SINR}$, thus contradicting the optimality of $\hat t$. Alternatively, if $\hat t > t $ then the solution set $ \{\mathbf w_k^F \}_{k=1   }^{G}$ can be scaled by a positive constant $ c = t /\hat t<1$. The new solution  $ \{c \mathbf w_k^F \}_{k=1   }^{G}$ respects the feasibility conditions of $\mathcal{Q}$ and provides a lower optimal value, i.e. $c\cdot r^*$, thus again contradicting the hypothesis. As a result, $\hat t = t\ \square$.

 \subsection{Semidefinite Relaxation}\label{sec: SDR}
 Problem  $\mathcal{Q}$ belongs in the general class of non-convex $\mathrm{QCQP}$s for which the $\mathrm{SDR}$ technique is proven to be a powerful and computationally efficient approximation technique \cite{Luo2010}. The relaxation is based on the observation that
$|\mathbf w_k^\dag \mathbf h_i|^2 = \mathbf w_k^\dag \mathbf h_i \mathbf h_i^\dag \mathbf w_k = {\mathrm{Tr}}(\mathbf w_k^\dag \mathbf h_i \mathbf h_i^\dag \mathbf w_k) = {\mathrm{Tr}}(\mathbf w_k \mathbf w_k^\dag \mathbf h_i \mathbf h_i^\dag )$. With the change of variables $\mathbf X_i = \mathbf w_i \mathbf w_i^\dag $,
 $\mathcal{Q}$ can be relaxed to
 $\mathcal{Q}_r$
\begin{empheq}[box=\fbox]{align}
\mathcal{Q}_r:\min_{r,\ \{\mathbf X_k \}_{k=1}^{G}}  &r& \notag\\
\mbox{subject to } & \frac{\mathrm{Tr}\left(\mathbf Q_i \mathbf X_k\right)}{\sum_{l\neq k }^G \mathrm{Tr}\left(\mathbf Q_l \mathbf X_k\right) +\sigma_i^2}\geq \gamma_i, \label{const: Q_r SINR}\\
&\forall i \in\mathcal{G}_k, k,l\in\{1\dots G\},\notag\\
 \text{and to} \ \ \ \ \ &\frac{1}{P_n} \left[\sum_{k=1}^G \mathbf X_{k}\right]_{nn} \leq  r   \label{const: Q_r Power}\\
 \text{and to}  \ \ \ \ \ &\ \mathbf X_k\succeq0,\label{const: Q_r SDF}
\ \forall n\in \{1\dots N_{t}\},\notag
 \end{empheq}
where $\mathbf Q_i = \mathbf h_i\mathbf h_i^\dag$, $r \in \mathbb{R}^{+}$ , while the constraint $\text{rank}(\mathbf X_i)  = 1$ is dropped. Now the relaxed  $\mathcal{Q}_r$ is convex, thus solvable to an arbitrary accuracy. {This relaxation can be  interpreted as a Lagrangian bi-dual of the original problem  \cite{convex_book}. }The weighted max-min fair optimization is  also relaxed as
\begin{empheq}[box=\fbox]{align}
\mathcal{F}_r:\max_{t, \ \{\mathbf w_k \}_{k=1}^{G}}&t& \notag\\
\mbox{subject to } & \frac{1}{\gamma_i}\frac{\mathrm{Tr}\left(\mathbf Q_i \mathbf X_k\right)}{\sum_{l\neq k }^G \mathrm{Tr}\left(\mathbf Q_l \mathbf X_k\right)+ \sigma_i^2}\geq t, \label{const: F_r SINR}\\
&\forall i \in\mathcal{G}_k, k,l\in\{1\dots G\}, \notag\\
\mbox{and to }\ \ \ \ &\left[\sum_{k=1}^G  \mathbf X_k \right]_{nn} \leq  P_n,\\
  &\forall n\in \{1\dots N_{t}\},\notag \\
 \mbox{and to }\ \ \ \ &  \mathbf X_k\succeq0,
\end{empheq}
{which, however, remains non-convex due to  \eqref{const: F_r  SINR}, as in detail explained in \cite{Karipidis2008}}. However, this obstacle can  be overcome by  the following observation.

\textit{ \textit{Claim 2}: Problems $\mathcal{F}_r$ and  $\mathcal{Q}_r$ are related as follows}
\begin{align}
1 = \mathcal{Q}_r\left(\mathcal{F}_{r}\left( \mathbf g, \mathbf p\right)\cdot\mathbf g, \mathbf p \right)\label{eq: equivalence 3}\\
t = \mathcal{F}_r\left(\mathbf g, \mathcal{Q}_{r}\left( t\cdot \mathbf g,\mathbf p\right)\cdot\mathbf p \right)\label{eq: equivalence 4}
 \end{align}
\textit{Proof:} Follows the steps of the proof of \textit{Claim 1} and is therefore omitted. $\square$
\subsection{Gaussian Randomization} \label{sec: Gaussian randomization}
Due to the NP-hardness of the multicast problem,   the relaxed problems do not necessarily yield unit rank matrices. Consequently, one can  apply a rank-1 approximation over $\mathbf X^{*}$.
Many types of rank-1 approximations are possible depending on the nature of the original problem.
The solution with the highest provable accuracy for the multicast case is given by the {Gaussian randomization method \cite{Luo2010}.} In more detail, let  $\mathbf X^{*}$ be a symmetric positive semidefinite solution of the relaxed problem. Then, a candidate solution to the original problem can be generated as a Gaussian random variable with zero mean and covariance equal to $\mathbf X^{*}$, i.e.  $\hat{\mathbf{w}} \backsim\mathbb{C}\mathbb{N}(0, \mathbf X^{*} )$.
After generating a predetermined number of candidate solutions, the one that yields the highest objective value of the original problem can be chosen. The accuracy of this approximate solution is measured by the distance of the approximate objective value and the optimal value of the relaxed problem and it increases with the predetermined number of  randomizations {\cite{Luo2010,Karipidis2008}}.
Nonetheless, an intermediate problem dependent step between generating a Gaussian instance with the statistics obtained from the relaxed solution and creating a feasible candidate instance of the original problem still remains, since the feasibility of the original problem is not yet guaranteed.

\subsection{Feasibility Power Control }\label{sec: power control}
After generating a random instance of a Gaussian variable with statistics defined by the relaxed problem, an additional step comes in play to guarantee the feasibility of the original problem. { In \cite{Sidiropoulos2006},   the feasibility of the candidate     solutions, as given by the Gaussian randomization, was guaranteed by a simple  power rescaling.} Nevertheless, since in the multigroup case  an interference scenario is dealt with,   a simple rescaling does not guarantee feasibility. Therefore, an additional  optimization step is proposed in \cite{Karipidis2008} to re-distribute the power amongst the candidate precoders. To  account for the inherently different $\mathrm{PAC}$s,  a novel power control problem with per antenna power constraints is  proposed. Given a set of  Gaussian instances,  $\{\mathbf{\hat w}_k\}_{k=1 }^G$, the \textit{Multigroup Multicast Per Antenna power Control} ($\mathrm{MMPAC}$) problem reads as
\begin{empheq}[box=\fbox]{align}
\mathcal{S^{\mathcal{F}}:} \max_{t, \ \{  p_{k} \}_{k=1}^{G}}  &t& \notag\\
\mbox{subject to } & \frac{1}{\gamma_i}\frac{|\mathbf{\hat  w}_k^\dag \mathbf{  h}_i|^2 p_{k}}{\sum_{l\neq k }^G |\mathbf{\hat  w}_l \mathbf{ h}_i |^2  p_{l} +\sigma_i^2}\geq t, \label{const: S F SINR}\\
&\forall i \in\mathcal{G}_k,k,l\in\{1\dots G\}\notag \\
\mbox{and  to}  \ \ \ \ \ & \left[\sum_{k=1}^G \mathbf{\hat  w}_k\mathbf{\hat  w}_k^\dag  p_k\right]_{nn} \leq   P_{n,} \label{eq: MMPAC PACs}\\
&\forall n\in \{1\dots N_{t}\},\notag
 \end{empheq}
with $\{  p_{k} \}_{k=1}^{G}\in\mathbb{R}^{+}$. Problem $\mathcal{S^{\mathcal{F}}}$ receives as input the $\mathrm{PAC}$s as well as the  $\mathrm{SINR }$ targets and returns the maximum scaled worst $\mathrm{SINR}$ $t^{*} = \mathcal{S}(\mathbf g, \mathbf p)$ and is also non-convex like $ \mathcal{F}$.
{{The difference of this problem compared to \cite{Karipidis2008} lies in \eqref{eq: MMPAC PACs}.
}
\textit{\\ Remark 2:} A very important observation is clear in the formulation of the power control problem. The optimization variable $\mathbf p $ is of size $G$, i.e. equal to the number of groups, while the power constraints are equal to the number of antennas, $N_t$. In each constraint, all the optimization variables contribute. This fact prohibits the total exploitation of the available power at the transmitter. Once at least one of the $N_t$ constraints is satisfied with equality and remaining power budget, then the rest can not be scaled up since this would lead to at least one constraint exceeding the maximum permitted value.}

\subsection{Bisection}\label{sec: bisec}

{The establishment of claims 1 and 2, allows for the application of the bisection method, as developed in \cite{Sidiropoulos2006,Karipidis2008}.} The solution of $
r^* = \mathcal{Q}_r \left ( \frac{L+U}{2}\mathbf g, \mathbf p\right)
$ is obtained by bisecting the interval $[L, U]$  as defined by the minimum and maximum $\mathrm{SINR}$ values. Since $t=(L+U)/2$ represents the $\mathrm{SINR}$
, it will always be positive or zero. Thus, $L = 0.$ Also, if the system was interference  free while all the users had the channel of the best user, then the maximum worst $\mathrm{SINR}$ would be attained, thus $U = \max_i\{P_{tot}\mathbf Q_i/\sigma_i \}. $ If $r^*<1$,   then the lower bound of the interval is updated with this value. Otherwise the value is assigned to the upper bound of the interval. Bisection is iteratively performed until an the interval size is reduced to a pre-specified value $\epsilon$.
This value needs to be dependent on the magnitude of $L \text{ and } U$ so that the accuracy of the solution is maintained regardless of the region of operation.
 After a finite number of iterations the optimal value of $\mathcal{F}_r$ is given as the resulting value for which $L \text{ and } U$ become almost identical. This procedure provides an accurate solution to the  non-convex $\mathcal{F}_r$.
Following this, for each and every solution $\{\mathbf{\hat w}_k\}_{k=1 }^G$, the power of the precoders needs to be controlled. 
{Consequently, problem $\mathcal{S}^F$ can be solved using the well established framework of  bisection \cite{convex_book} over its convex equivalent problem, which reads as}
\begin{empheq}[box=\fbox]{align}
\mathcal{S^{\mathcal{Q}}:} \min_{r, \ \{ p_k \}_{k=1}^{G}}  &r& \notag\\
\mbox{subject to } & \frac{|\mathbf{\hat  w}_k^\dag \mathbf h_i|^2 p_{k}}{\sum_{l\neq k }^G | \mathbf{\hat  w}_l^\dag\mathbf h_i|^2  p_{l} +\sigma_i^2}\geq \mathbf \gamma_i, \label{const: F SINR}\\
&\forall i \in\mathcal{G}_k, k,l\in\{1\dots G\},\notag \\
\mbox{and  to}  \ \ \ \ \ & \frac{1}{  P_{n}} \left[\sum_{k=1}^G  \mathbf{\hat  w}_k\mathbf{\hat  w}_k^\dag  p_k\right]_{nn} \leq  r, \\
&\forall n\in \{1\dots N_{t}\},\notag
 \end{empheq}
{ Problem $\mathcal{S^\mathcal{Q}}$ is an instance of a linear programming (LP) problem.} 

{\textit{Remark
3}:
For  completeness,
the
possible
reformulation of the non-convex problem
$\mathcal{S^{\mathcal{F}}} $ into the following geometric problem ($\mathrm{GP}$)   is considered, thus surpassing the need for bisection:}
\begin{empheq}[box=\fbox]{align}
\mathcal{S^{\mathcal{F}}_{\mathcal{GP}}}:& \min_{t, \ \{  p_{k} \}_{k=1}^{G}}  t^{-1}& \notag\\
\mbox{s. t. }& {\sum_{l\neq k }^G |\mathbf{\hat  w}_l \mathbf{ h}_i |^2  p_{l} +\sigma_i^2}\leq \frac{t^{-1}}{\gamma_i}|\mathbf{\hat  w}_k^\dag \mathbf{  h}_i|^2 p_{k}, \label{const: S F SINR}\\
&\forall i \in\mathcal{G}_k,k,l\in\{1\dots G\}\notag \\
\mbox{and  to} &  \left[\sum_{k=1}^G \mathbf{\hat  w}_k\mathbf{\hat  w}_k^\dag  p_k\right]_{nn} \leq   P_{n,} \label{eq: MMPAC PACs GP}
\forall n\in \{1\dots N_{t}\},\notag
 \end{empheq}
\subsection{Complexity}\label{sec: complexity}
 An important discussion involves the complexity of the employed techniques  to approximate a solution of  the highly complex, NP-hard  multigroup multicast problem under $\mathrm{PAC}$s.   Focusing on the proposed algorithm (cf. Alg. 1), the main complexity burden originates from the solution of a $\mathrm{SDP}$. The present work relies on  the CVX tool \cite{convex_book} which calls  numerical  solvers such as SeDuMi to solve  semi-definite programs. The complexity of the $\mathrm{SDR}$  technique has been  exhaustively discussed in  \cite{Luo2010} and the references therein.    To calculate the total worst case complexity of the  solution proposed in the present work, the following are considered.

Initially, a  bisection search is performed over  $\mathcal{Q}_r$ to  obtain the relaxed solution.  This bisection runs for $N_{iter} = \lceil\log_2\left(U_{1}-L_{1}\right)/\epsilon_{1}\rceil$ where $\epsilon_1 $ is the desired accuracy of the search.  Typically   $\epsilon_1 $ needs to be at least three orders of magnitude below the magnitudes of $U_{1}, L_{1}$ for sufficient accuracy. In each iteration of the bisection search, problem $\mathcal{Q}_r$ is solved.  This $\mathrm{SDP}$ has $G$ matrix variables of $N_t\times N_t$ dimensions and $N_{u}+N_t$ linear constraints.   The interior point methods employed to solve this $\mathrm{SDP}$  require at most { $\mathcal{O}\left(\sqrt{GN_t}\log(1/\epsilon) \right)$ } iterations, where $\epsilon $ is the desired numerical accuracy of the solver. Moreover, in each iteration not more than $\mathcal{O}({G^3 N_t^6 +GN_{t}^{3} +N_{u}GN_t^2})$ arithmetic operations will be performed.  The increase in complexity stems from increasing the number of constraints, i.e. $N_t+N_{u}$  constraints are  considered instead of only $N_{u}$ as in \cite{Karipidis2008}. However, this increase is not significant, since  the order of the polynomial with respect to the number of transmit antennas is not increased.   The solver used also exploits the specific structure of matrices hence the actual running time is reduced. Next,  a fixed number of Gaussian random instances with covariance given by the previous solution are generated. {The complexity burden of this step is given by the following considerations. For each randomization, a second bisection search is performed this time over the $\mathrm{LP}$ $\mathcal{S}^Q$. An $\epsilon-$optimal solution of this problem can be generated with a worst case complexity of  $\mathcal O(G^{3.5}\log(1/\epsilon))$ \cite{ye2011interior} . The second bisection runs for $N_{iter} = \lceil\log_2\left(U_{2}-L_{2}\right)/\epsilon_{2}\rceil$ iterations, which are significantly reduced   since the upper bound $U_{2}$ is now the optimal value of the relaxed problem.
 Moreover, the Gaussian randomization  is executed for a fixed number of iterations. The accuracy of the solution increases with the number of randomizations \cite{Karipidis2008,Sidiropoulos2006,Luo2010}. Finally, the complexity burden can be further reduced by the reformulation of the non-convex
 $ \mathcal{S}^{\mathcal{F}} $
 into the $\mathrm{GP}$, $\mathcal{S^{\mathcal{F}}_{\mathcal{GP}}}$ which is efficiently solved by successive approximations of primal-dual interior point numerical methods \cite{convex_book}. Thus the need for the second bisection can be surpassed.  }
\begin{algorithm}
 \SetAlgoLined 
 \KwIn{$ N_{rand}, \mathbf p, \mathbf g, \mathbf Q_i, \sigma_i^2 \ \forall i \in\{1\dots G\} $ }
 \KwOut{   $ \{\mathbf w_{k}^{opt}\}_{k=1}^G$, $t^*_{opt}$ of $\mathcal{F}$ $ \{\mathbf w_{k}^{out}\}_{k=1}^G$   $t^*_{out}$  }
 \Begin{
 \textbf{\textit{\textit{\uline{Step 1:}}}} Solve $t_{opt} = \mathcal{F}_r\left(\mathbf g, \mathbf p \right)$ by bisecting  $\mathcal{Q}_r\left(\frac{L+U}{2}\mathbf g, \mathbf p \right)$, (see Sec. \ref{sec: bisec}).
Let  the associated point be $\{\mathbf w_{k}^{opt}\}_{k=1}^G$.\\
 \eIf{$\mathrm{rank}(\mathbf X_{k}^{opt} )= 1, \forall \ k \in\{1\dots G\} $}
{$ \{\mathbf w_{k}^{out}\}_{k=1}^G$ is given by $\lambda_{max}(\mathbf X^{opt})$.  }
{
\textbf{\textit{\uline{Step 2:}}}
: Generate $N_{rand}$  precoding vectors
$\{ \mathbf{\hat{w}}_{k}\}_{k=1}^G$, (see Sec. \ref{sec: Gaussian randomization} ).
$t^{*}_{(0)}= 0$\;
\For{$i=1\dots N_{rand}$}{
\textbf{\textit{\uline{Step 3:}}}
Solve $\mathcal{S}^{\mathcal{F}}\left(\mathbf g, \mathbf p \right)$ by bisecting $\mathcal{S}^{\mathcal{Q}}\left(\frac{L+U}{2}\mathbf g, \mathbf p \right)$ to get a $\{\mathbf w_{k}^{can}\}_{k=1}^G$ with $t_{(i)}^*$.\\
 \If {$t_{(i)}^* > t_{(i-1)}$ }{ $t^{*}_{out}=t^*_{(i)}, \{\mathbf w_{k}^{out}\}_{k=1}^G = \{\mathbf w_{k}^{can}\}_{k=1}^G$}
}
}
}
 \caption{ Fair multigroup multicasting under $\mathrm{PAC}$s.}
\end{algorithm}

\section{ Performance Evaluation \& Applications} \label{sec: performance}
\subsection{Multigroup multicasting over Rayleigh Channels}\label{sec: performance rayleigh}
The performance of linear multicast multigroup beamforming under per antenna power constraints is examined for a system with $N_t = 5$ transmit antennas, $G = 2 $ groups and  $N_u = 4  $ users. Rayleigh fading is considered, thus the channels are generated as Gaussian complex variable instances with unit variance and zero mean. For every channel instance, the  approximate solutions of the max-min fair $\mathrm{SPC}$   and the proposed $\mathrm{PAC}$ problems are evaluated using $N_{rand} = 100$ Gaussian randomizations\cite{Karipidis2008}.
The results are averaged over one hundred channel realizations, while the noise variance is normalized to one for all receivers and all $\mathrm{SINR}$ targets are  assumed equal to one.

 {The achievable  minimum rate
 is plotted for the $\mathrm{SPC}$ and the $\mathrm{PAC  }$ optimization  in Fig. \ref{fig:  power}
 with respect to the total transmit power in dBWs. Noise   is assumed normalised to one.} For fair comparison, the total power constraint $P_{tot}$~[Watts] is equally distributed amongst the transmit antennas when $\mathrm{PAC}$s  are considered, hence each antenna can radiate at most $P_{tot}/N_t $~[Watts].  The  accuracy of the approximate solutions for both problems, {given by comparing the actual solution to the relaxed upper bound \cite{Sidiropoulos2006,Karipidis2008},}     is clear across a wide range of $\mathrm{SNR}$. Nevertheless, the accuracy due to the $\mathrm{PAC  }$s is slightly reduced. This accuracy degradation  is intuitively justified. A Gaussian randomization instance is less likely to approach the optimal point when the number of constraints is increased while the same number of Gaussian randomizations are performed ($N_\mathrm{rand} = 100$).
{Towards quantifying the gains of the proposed solution, the performance of the $\mathrm{SPC  }$ solution re-scaled to respect the $\mathrm{PAC  }$s is also included in Fig. 1. Re-scaling is achieved by multiplying each line of the precoding matrix with the square root of the inverse level of power over satisfaction of the corresponding antenna. In Fig. 1 it is clear that more than 1 dB of gain can be obtained by the proposed method over the suboptimal re-scaling approach.}

A significant issue for the $\mathrm{SDR}$ techniques in multicast applications is the tightness of the approximate solution  versus an increasing number of receivers per multicast. In the extreme case of one user per group, it was proven in\cite{Bengtsson2001} that the relaxation provides an optimal solution. Thus the solution is no longer approximate but exact. However, the increasing number of users per group degrades the solution, as depicted in Fig. \ref{fig:  users} for both problems. It is especially noticed that the $\mathrm{PAC}$ system suffers more than  the $\mathrm{SPC}$ of \cite{Karipidis2008} as the number of users per multicast group increases.  An attempt to solve this inaccuracy, but only under sum power constraints, is presented in \cite{Pesavento2012b}.
\begin{figure}[h]
\centering
 \includegraphics[width=0.8\columnwidth]{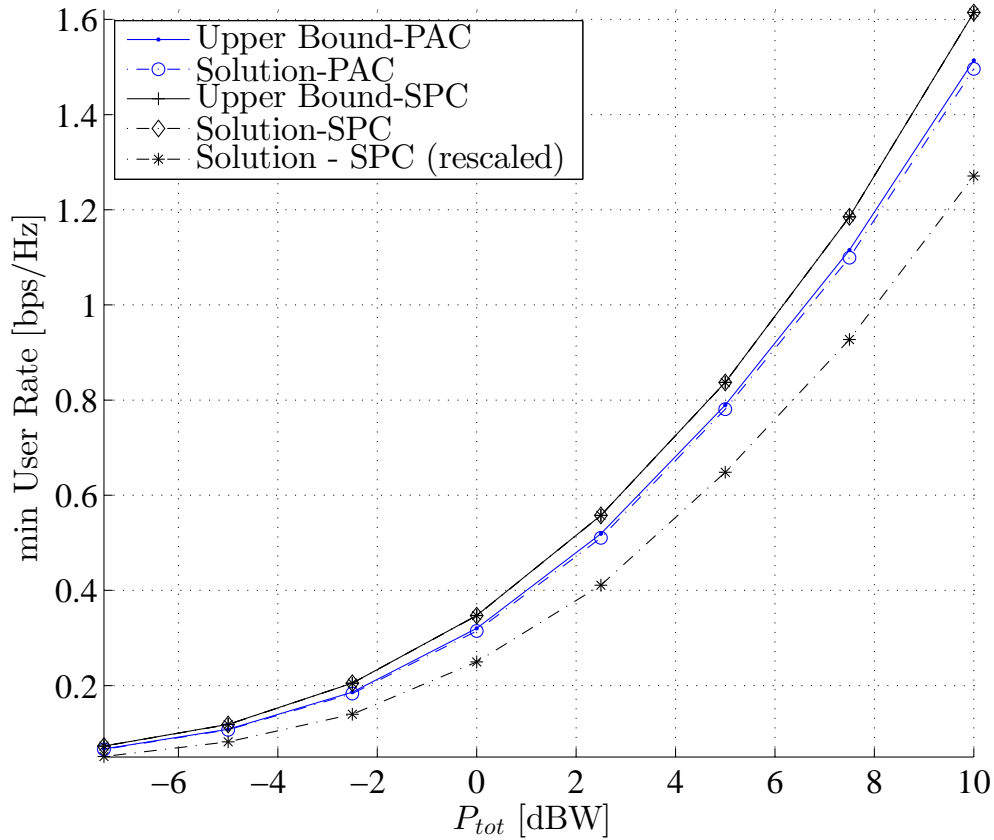}\\
 \caption{Minimum user rate with $\mathrm{SPC}$ and $\mathrm{PAC}$.
 Results for $N_{u} = 4$ users, $N_t = 5$ antennas, $L= 2$ groups and   $\rho = 2$ users per group. 
}\label{fig:  power}
 \end{figure}
\begin{figure}[h]
\centering
\includegraphics[width=0.8\columnwidth]{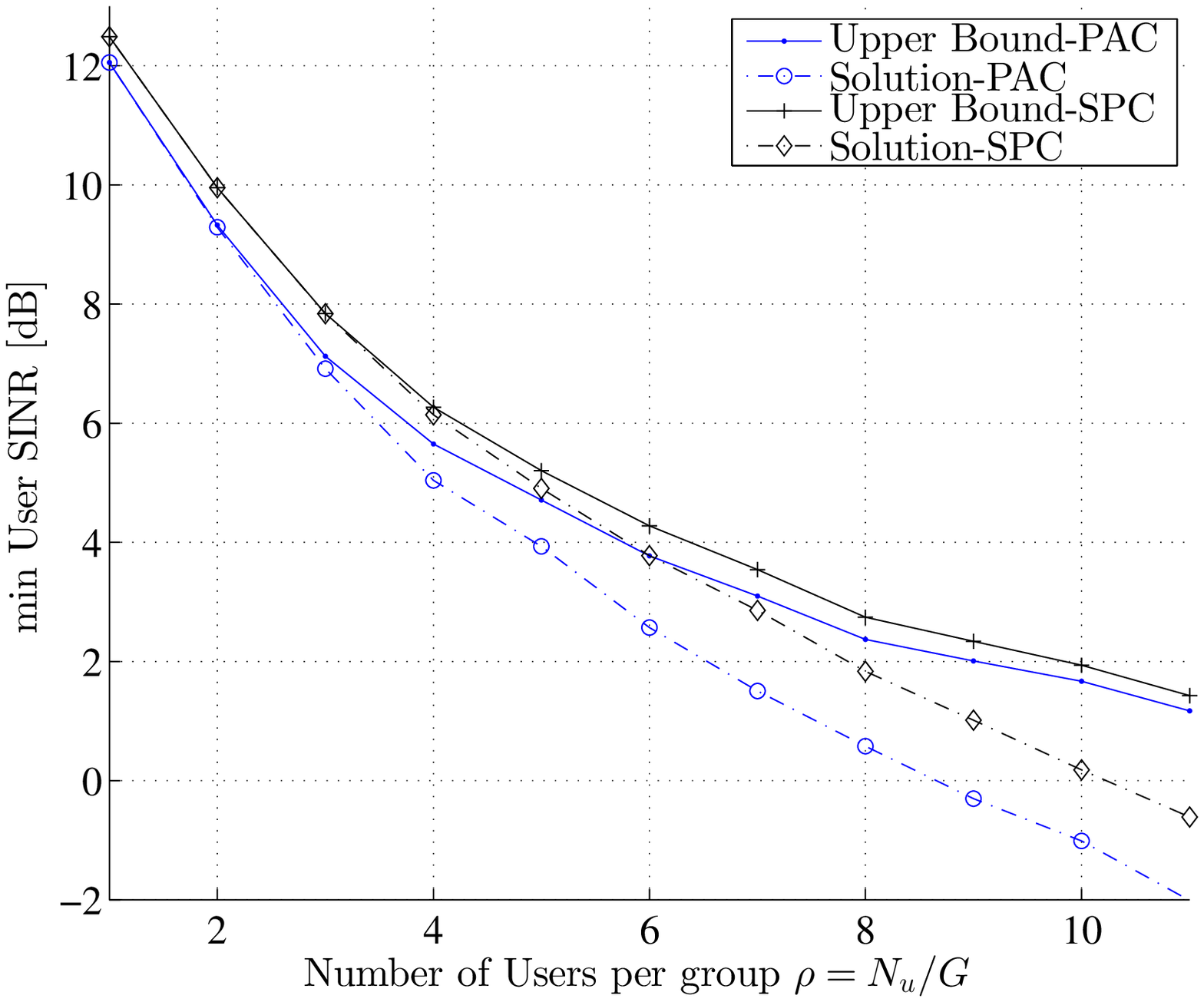}\\
\caption{Minimum $\mathrm{SINR}$ with $\mathrm{SPC}$ and $\mathrm{PAC}$  versus an increasing  ratio of users per group $ \rho = N_u / G  $,  for  ${P}_{tot} = 10$~dBW. 
}\label{fig:  users}
\end{figure}
\subsection{Power Consumption in DAS }\label{sec: power consumption}
The main difference between the $\mathrm{SPC}$ and the $\mathrm{PAC}$  optimization problems is the utilization of the available on board power in each system architecture.
In \cite{Karipidis2008},  the sum power constraint is always satisfied with equality, since any remaining power budget can be equally distributed to the precoding vectors and the solution is further maximized.
On the contrary, the $\mathrm{PAC }$ system includes $N_t$  constraints which are coupled via the precoders.   According to the relation between $\mathcal{F}$  and $\mathcal{Q}$, i.e. \eqref{eq: equivalence 1}, the ratio of transmitted power over the power constraint (i.e. $r$) is one. Since this ratio applies for at least one of the $N_t$ power constraints, if one  is met with equality and the remaining $N_t - 1$ are not, then no more power can be allocated to the precoders.
Let us assume a  channel matrix with one compromised transmit antenna, i.e. $\mathbf{H}=$
\begin{align}
 \begin{bmatrix}\notag
2.94\angle41^\circ & 11\angle{-25}^\circ & 4.4\angle{50}^\circ &6.6\angle{-4}^\circ \\
 13.2 \angle{-150}^\circ & 4.8\angle{14}^\circ & 15.2\angle{-7}^\circ &4.8\angle{-37}^\circ \\
 12 \angle{-155}^\circ&1.5\angle{163}^\circ & 13.5\angle{-105}^\circ &3.9\angle{-46}^\circ \\
 0.02 \angle{-53}^\circ & 0.03\angle{-66}^\circ & 0.03\angle{120}^\circ &0.03\angle{-129}^\circ \\
 5.66 \angle{137}^\circ &   9.2\angle{49}^\circ & 13\angle{-175}^\circ & 2.45\angle{126}^\circ  \\
\end{bmatrix} ^\text T,
 \end{align}
where $4$ users, divided into $2$ groups,  are served by $5$ antennas. One of the antennas (the $4$-th antenna) has severely degraded gains towards all users. This practical case can  appear in a $\mathrm{DAS }$ where the physical separation of the transmit antennas not only imposes per antenna constraints but can also justify highly unbalanced channel conditions around the environment the antennas.
The power utilization of the solution of the optimization for each of the two problems is defined as the total transmitted power over the total available power $P_{tot}$, that is $P_u =\left(\sum_{k=1}^ G \mathbf w_k{^\dag} \mathbf w_k\right)/{P_{tot}}$,
 and is  plotted versus an increasing power budget in Fig. \ref{fig: power consumption}. It is clear that in the low power regime the available power is not fully utilized. As the available power increases, however, the power consumption of the $\mathrm{PAC}$ increases. This result is in accordance with the optimality of equal power allocation in the high power regime and renders  the $\mathrm{PAC}$ formulation relevant for power limited systems.
Further insights for this $\mathrm{PAC}$ system are given in Fig. \ref{fig: power bar}, where   the power utilization of each antenna is shown,  for different total power budgets.
Interpreting these  results, it can be concluded that the $\mathrm{PAC }$ problem is highly relevant for power-over-noise limited systems. Otherwise, in the high  power regime, the solution of the $\mathrm{SPC }$ problem with less constraints could be also used as an accurate approximation. 
 \begin{figure}[h]
 \centering
 \includegraphics[width=0.8\columnwidth]{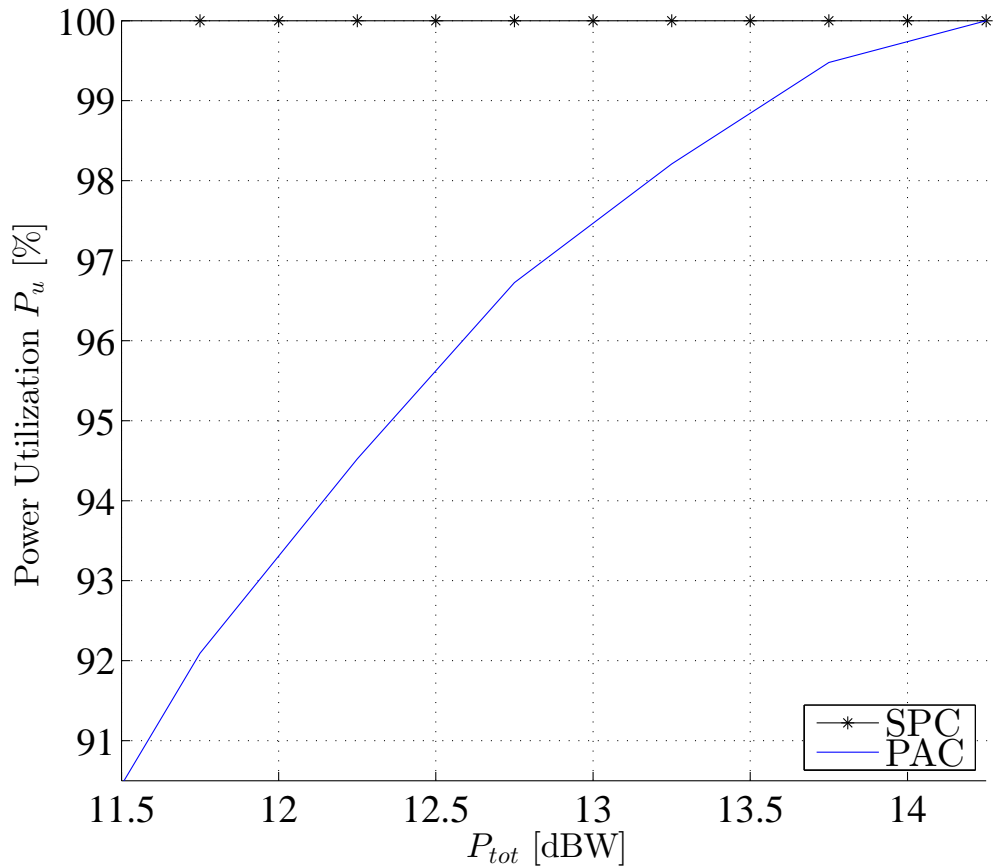}\\
 \caption{Total power consumption of a   $\mathrm{PAC}$  system versus available power.}\label{fig:  power consumption}
 \end{figure}
 \begin{figure}[h]
 \centering
 \includegraphics[width=0.8\columnwidth]{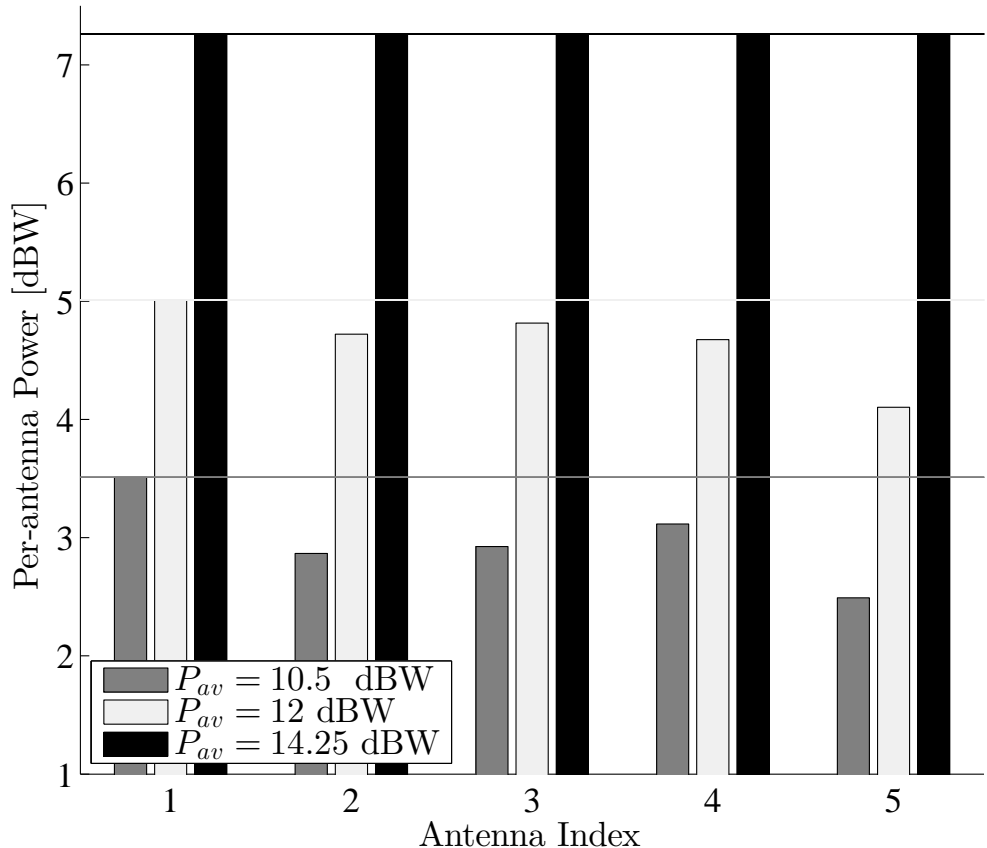}\\
 \caption{Per-antenna  consumption  in a  $\mathrm{PAC}$ system  versus   transmit power. 
}\label{fig:  power bar}
\end{figure}
\subsection{Weighted Fairness Paradigm}
To the end of establishing the importance of the weighted optimization,  a simple paradigm is elaborated herein. Under the practical assumption of a modulation constrained system, the weighted fair design can be exploited for  rate allocation towards increasing the total system throughput.
 More specifically, the considered system employs adaptive modulation and allocates binary phase shift keying ($\mathrm{BPSK}$) modulation  if the minimum $\mathrm{SINR_i}$  in the $k$-th group is less than the ratio for which the maximum modulation constrained spectral efficiency is achieved. This ratio is simply given by $ \log_2 M$, where $M $ is the modulation order. Hence for $\mathrm{BPSK}$, $\gamma_2 = 0$~dB, and so forth. If for some group $k$,  $\min_i{\mathrm{SINR_i}} \geq \gamma_{ \mathrm{2}},\ \forall i \in\mathcal{G}_k, $ then quaternary phase shift keying ($\mathrm{QPSK}$) is used  for all users in the group. Forward error correction is not assumed.  Let there be a two antenna transmitter that serves four users   grouped into two groups. The considered channel matrix reads as
 \begin{align}\mathbf{H }=
 \begin{bmatrix}\notag
 0.2 \angle106^\circ & 90\angle{-69}^\circ & 0.5\angle{-99}^\circ &0.5\angle{61}^\circ \\
 0.8 \angle111^\circ &   120 \angle{-112}^\circ & 1\angle{127}^\circ & 1.5\angle{49}^\circ  \\
\end{bmatrix} ^\text T.
 \end{align}
The attributes of the specific channel matrix depict one possible instance of the system where one user with a good channel state (i.e. user two) is in the same group with a jeopardized user, namely user one. On the other hand, the second group contains relatively balanced users in terms of channel conditions. For an un-weighted optimization (i.e. $\mathbf g = [1\ 1\ 1\ 1]$) the  spectral efficiency of each user is shown in  Fig. \ref{fig: spf bar}. Baring in mind  that each user is constrained by the minimum group rate, the actual rate at which  all users will receive data is 0.52 [bps/Hz]. Both groups  achieve the same spectral efficiency since the minimum $\mathrm{SINR}$s and hence the minimum rates are balanced between the groups.   Subsequently, a modulation constrained  multicast transmitter  will  employ $\mathrm{BPSK}$ for all users.
 By heuristically choosing the constraint vector  to be  $\mathbf g = [1\ 1\ 5.3\ 5.3]$ each user rate is modified. As depicted in Fig. \ref{fig: spf bar} both users in the second group are achieving adequate   $\mathrm{SINR}$ to support a higher order modulation. This gain is achieved at the expense of the rates of the users of the first group.
Following this paradigm, the weight optimization can lead to an improved modulation assignment and thus higher throughput in practical systems. Hence, the weighted formulation offers the substantial degrees of freedom to maximize the total throughput of a modulation constrained multicast system by properly allocating the rates amongst the groups.
 \begin{figure}[h]
 \centering
 \includegraphics[width=0.8\columnwidth]{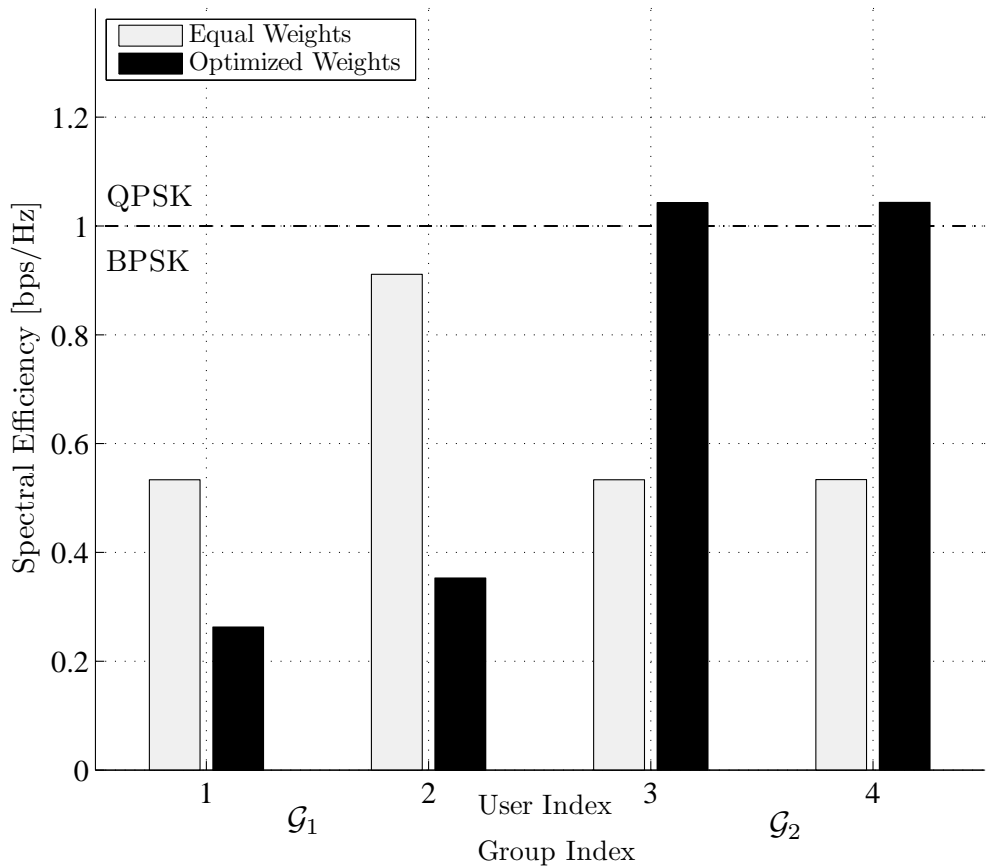}\\
  \caption{Modulation constrained paradigm.}
\label{fig: spf bar}
\end{figure}
\subsection{{Uniform Linear Arrays}}
To the end of investigating the sensitivity of the proposed algorithm with respect to the angular separation of co-group users, a uniform linear array ($\mathrm{ULA}$) transmitter is considered. Assuming far-field, line-of-sight conditions, the user  channels  can be  modeled using Vandermonde matrices. For this important special case, the $\mathrm{SPC}$ multicast multigroup problem was reformulated into a convex optimization problem and solved in \cite{Karipidis2006,Karipidis2007}. These results where motivated by the observation that in  $\mathrm{ULA}$ scenarios, the relaxation consistently yields rank one solutions. Thus, for such cases, the $\mathrm{SDR}$ is essentially optimal \cite{Sidiropoulos2006}. The fact that the $\mathrm{SDR}$ of the sum power minimization problem   is tight for Vandermonde channels was established in \cite{Karipidis2007}.
Let us consider a $\mathrm{ULA}$ serving $4$ users allocated to $2$ distinct groups.  In Fig. \ref{fig: ULA pos}, its radiation pattern for co-group angular separation $\theta_a = 35^\circ$  is plotted. The symmetricity due to the inherent ambiguity of the $\mathrm{ULA}$ is apparent. Clearly, the multigroup multicast beamforming optimizes the lobes  to reduce interferences between the two groups. The $\mathrm{SPC}$ solution, re-scaled to respect the $\mathrm{PAC}$s are also included in Fig. \ref{fig: ULA pos}. The superiority of the proposed solution is apparent.

In Fig. \ref{fig: ULA sr}, the performance in terms of minimum user rate over the area with respect to an increasing angular separation is investigated. When co-group users are collocated, i.e. $\theta_a = 0^\circ$, the highest performance is attained.  As the separation increases, the performance is reduced reaching the minimum when users from different groups are placed in the same position, i.e. $\theta_a = 45^\circ$. In Fig. \ref{fig: ULA sr}, the tightness of the relaxation for the  $\mathrm{SPC}$ problem \cite{Karipidis2007} is clear. However, the same does not apply for the proposed $\mathrm{PAC}$. As co-group  channels tend to become  orthogonal, the approximation becomes less tight. Nevertheless, $N_\mathrm{rand} = 200$ randomizations are sufficient to maintain the solution above the re-scaled $\mathrm{SPC}$, as shown in Fig.  \ref{fig: ULA sr}. Consequently, the proposed solution outperforms a re-scaled  to respect the per-antenna constraints, $\mathrm{SPC}$ solution, over the span of the angular separations.

\textit{Remark 4:} The semidefinite relaxation of the per-antenna power minimization problem in $\mathrm{ULA}$ transmitters is not always tight.

For every optimum high rank set of matrices $\{\mathbf { X}_k ^{opt}\}_{k=1}^G$, there exists a set of rank one positive semidefinite matrices  $\{\mathbf {\bar X}_k^{opt}\}_{k=1}^G$, i.e. $\mathrm{rank}(\mathbf {\bar X}_k ^{opt})=1, \forall k \in \{1\dots G\}  $,  which is equivalent with respect to the power received at each user, i.e $\mathrm{Tr}(\mathbf { X}_k ^{opt}\mathbf { Q}_i)=\mathrm{Tr}(\mathbf {\bar X}_k ^{opt}\mathbf { Q}_i) , \forall i \in\mathcal{G}_k, k,l\in\{1\dots G\}$. This result is based on the Riesz-F\'ejer theorem on real valued complex trigonometric polynomials \cite{Karipidis2007}. Therefore, the Vandermonde channels impose a specific structure to the $\mathrm{SPC}$ solution that allows for a convex reformulation.
The difference in the case tackle herein lies in the $N_t$  $\mathrm {PAC}$s, i.e. $\left[\sum_{k=1}^G  \mathbf X_k \right]_{nn} \leq  P_n, \forall n\in \{1\dots N_{t}\}$, in which the channel structure is not involved.    Thus, a rank-1 matrix is equivalent in terms of per user received power \cite{Karipidis2007} but not necessarily in terms of per-antenna consumed power, as shown herein.
 \begin{figure}[h]
 \centering
 \includegraphics[width=0.8\columnwidth]{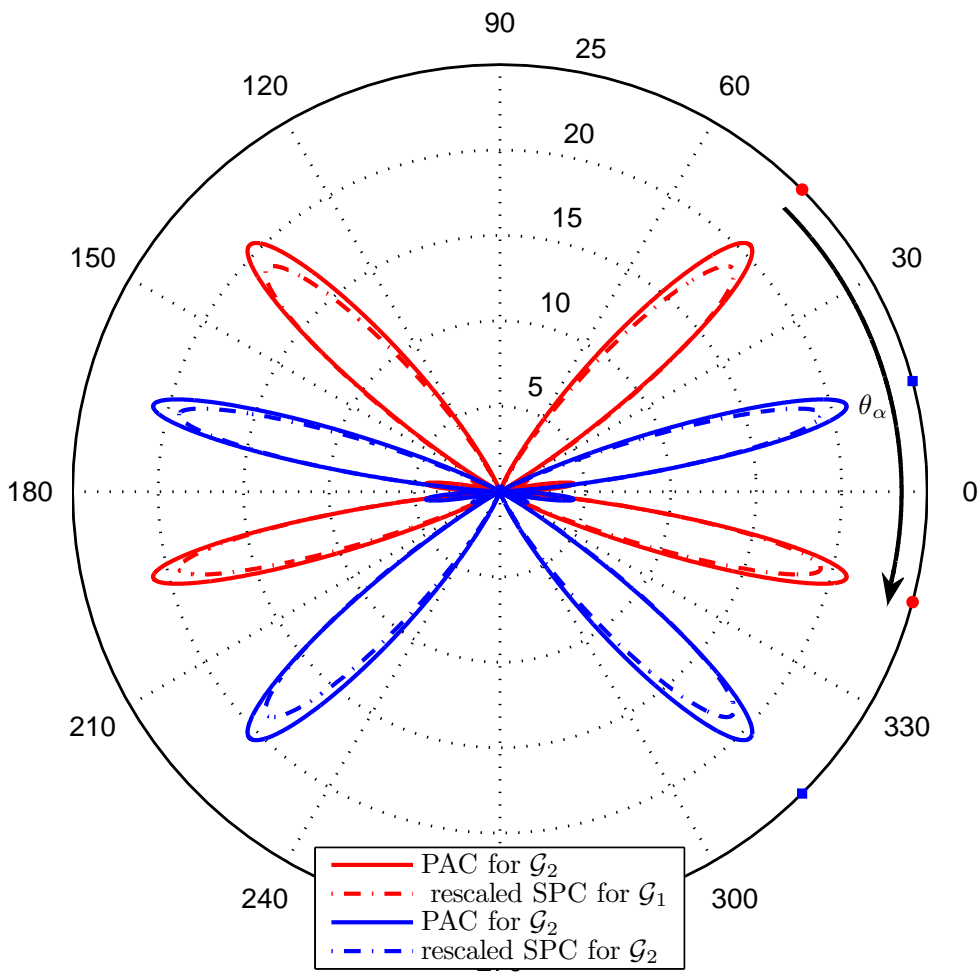}\\
  \caption{ULA beampattern for $\mathrm{PAC}$ and re-scaled $\mathrm{SPC}$ solutions.}
\label{fig: ULA pos}
\end{figure}
 \begin{figure}[h]
 \centering
 \includegraphics[width=0.8\columnwidth]{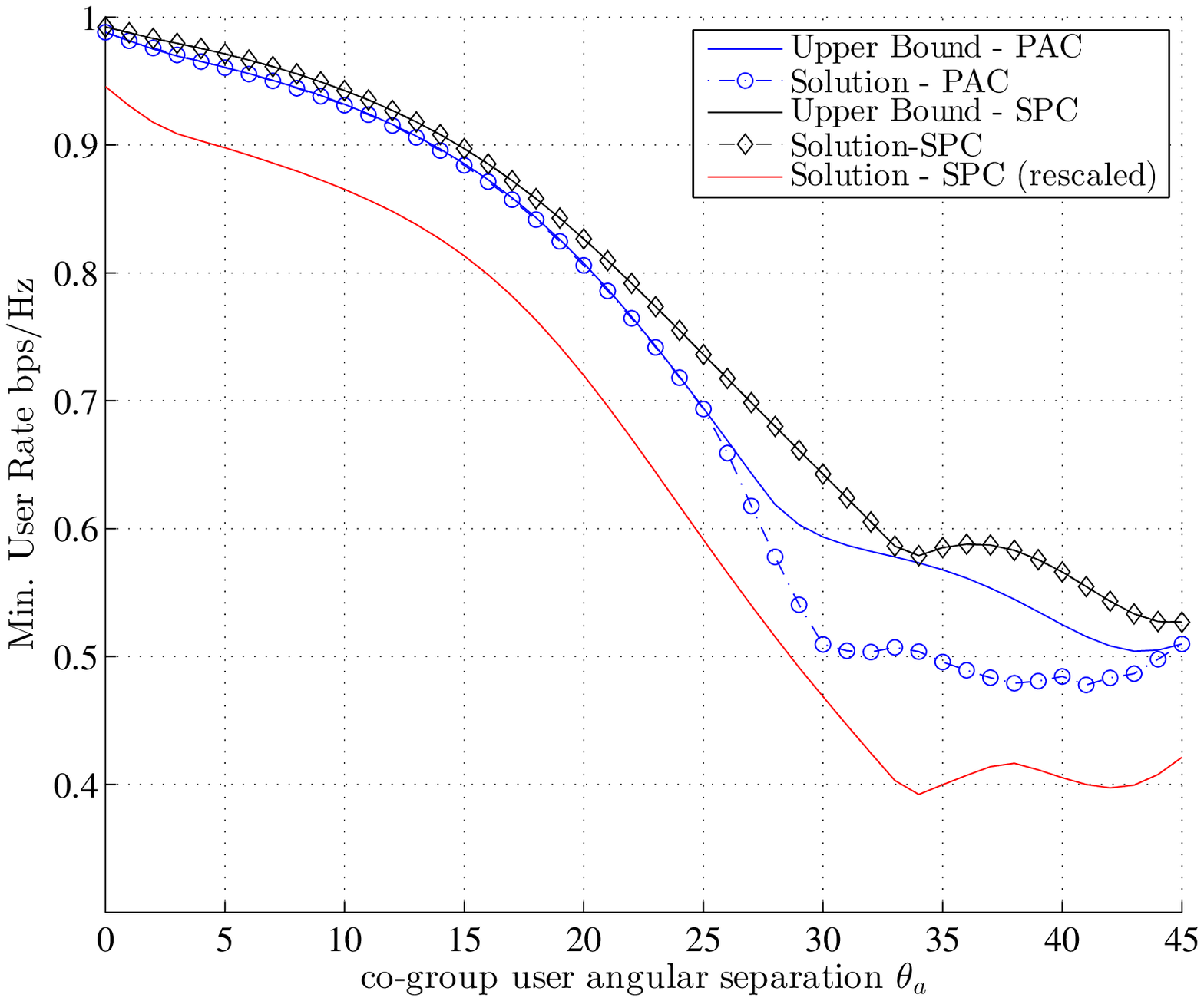}\\
  \caption{ $\mathrm{ULA}$ performance for increasing co-group user angular separation.}
\label{fig: ULA sr}
\end{figure}
\subsection{{Robust Design under $\mathrm{PAC}$s}}\label{sec: robustness}
 When beamforming under uncertainty is considered, three different designs can be realized\cite{Gershman2010}. Namely,  the probabilistic design, where acceptable performance is guaranteed for some percentage of time, the expectation based design that requires knowledge of the second order channel statistics but cannot guarantee any outage performance and the worst-case design. The latter approach guarantees a minimum $\mathrm{QoS}$ requirement  for any error realization.

Focusing on a worst-case design, let us assume an elliptically bounded error vector. In this context, the actual channel is given as $\mathbf h_{i} = \bar{ \mathbf h}_{i} +\mathbf e_{i}$  where $\bar{\mathbf h_{i}}$ is the channel available at the transmitter and $\mathbf e_{i}$ is an error vector  bounded by $\mathbf e_{i}^\dag \mathbf C_i \mathbf e_{i}\leq 1$. The hermitian positive definite matrix $\mathbf C_{i}$ defines the shape and size of the ellipsoidal bound. For  $\mathbf C_{i} = 1/\sigma_\epsilon^2\mathbf I_{N_t}$, then  $||\mathbf e_i||^2_2\leq \mathbf \sigma_\epsilon^2$ and  the error remains in a spherical region of radius $ \mathbf \sigma_\epsilon$ \cite{Shenouda2007}. This spherical error model is mostly relevant when the feedback quantization error of a uniform  quantizer at the receiver is considered \cite{Jindal2004}.
The proposed design is formulated as  \begin{empheq}[box=\fbox]{align}
\mathcal{F_{RB}:}&\max_{\  t, \ \{\mathbf w_k \}_{k=1}^{G}}  t\notag\\
\mbox{s. t. }  &\frac{1}{\gamma_i}\frac{|\mathbf w_k^\dag \left(\bar{ \mathbf h}_{i} +\mathbf e_{i}\right)|^2}{\sum_{l\neq k }^G |\mathbf w_l^\dag\left(\bar{ \mathbf h}_{i} +\mathbf e_{i}\right)|^2+\sigma_i^2 }\geq t, \label{const: FRB SINR}\\
&\forall i \in\mathcal{G}_k, k,l\in\{1\dots G\},\notag\\
 \text{and to }&  \left[\sum_{k=1}^G  \mathbf w_k\mathbf w_k^\dag  \right]_{nn}  \leq P_n, \forall n\in \{1\dots N_{t}\}\label{const: FRB PAC},
 \end{empheq}
and involves the channel imperfections  only in the $\mathrm{SINR}$ constraints. The novelty of $\mathcal{F_{RB}}$ over existing robust multicast formulations lies in \eqref{const: FRB PAC}.
The $\mathrm{SINR}$ constraints of $\mathcal{F_{RB}}$, i.e. \eqref{const: FRB SINR}, are over all possible error realizations and cannot be handled.  However, by applying the S-lemma \cite{convex_book},  the error vector in   \eqref{const: FRB SINR} can be eliminated.  This procedure is analytically described in  \cite{Chen2012}. Thus, $\mathcal{F_{RB}}$ can be converted to a $\mathrm{SDP}$ and solved efficiently using the methods described in Sec. \ref{sec: problem}.
The performance gain  of the proposed  robust design for a $\mathrm{ULA}$ with $N_t = 2$ transmit antennas, serving $N_{u}=6$ users is given in Fig. \ref{fig: robust SR config}, versus an increasing error radius $ \mathbf \sigma_\epsilon$, for different user per group   configurations,  $\rho$. These results exhibit the significant gains of the proposed technique as the error and   the group sizes increase.
 \begin{figure}[h]
 \centering
 \includegraphics[width=0.8\columnwidth]{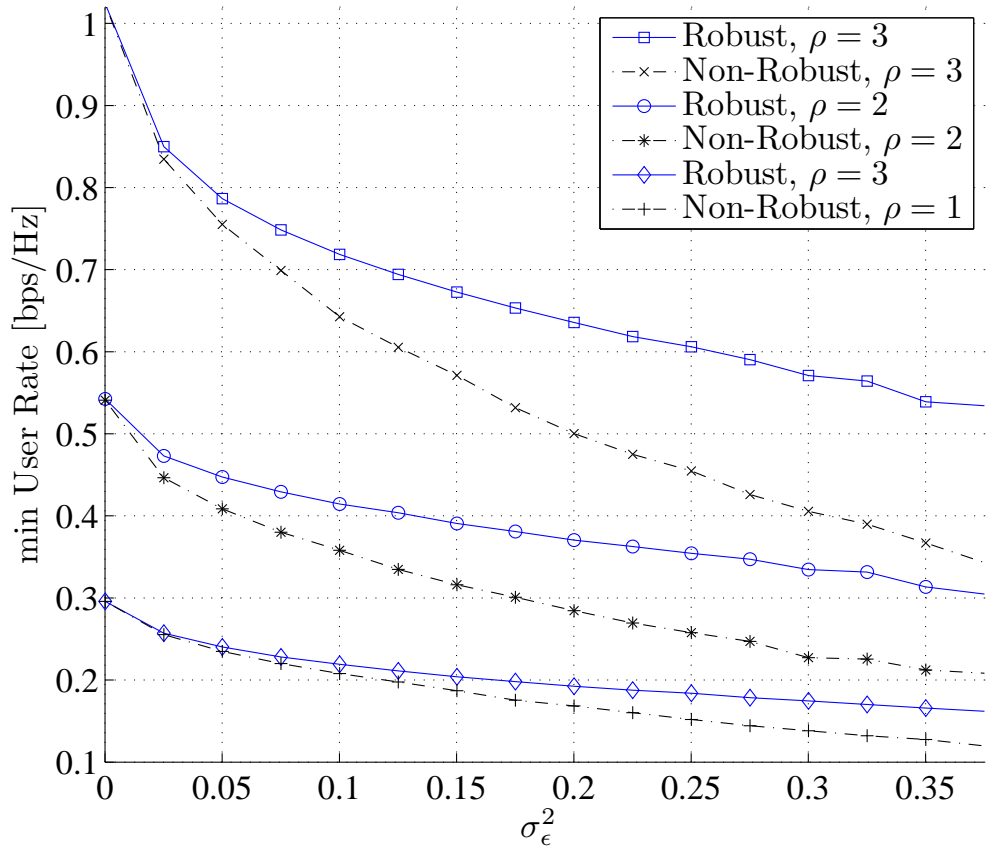}\\
  \caption{Robust performance for various user per group configurations.
}
\label{fig: robust SR config}
\end{figure}

 { To establish the importance of the novel formulation, the performance in terms of  minimum user rate   over 1000 error realizations is given in Fig. \ref{fig: robust SR}, versus a wide range of the error radius $ \mathbf \sigma_\epsilon$ for the proposed $\mathcal{F_{RB}}$ as well as the existing $\mathrm{SPC}$ solutions re-scaled to respect the per-antenna constrains. For this figure, a $\mathrm{ULA}$ with     $N_{t} = 3 $ transmit antennas is considered,  serving $N_{u}=6$ users partitioned into $L=2$ multicast groups.  The co-group angular separation is $\theta_a = 10^\circ$ and the number of Gaussian randomizations chosen is $N_\mathrm{rand} = 200$ and $N_\mathrm{rand} = 1000$ for the high and low precision curves respectively.  According to Fig. \ref{fig: robust SR},  the proposed robust $\mathrm{PAC}$ formulation (i.e. $\mathcal{F_{RB}}$) outperforms existing solutions, in a per-antenna power constrained setting, for a wide range of channel error radius.
 However, as the error radius increases, a slight performance degradation is noted, especially for the low precision results.
To further investigate on this result, the following remark is given.}

{\textit {Remark 5}: The semidefinite relaxation of robust multigroup multicasting under $\mathrm{PAC}$s yields non rank-1 solutions with higher probability as the channel errors increase.  }

{The accuracy of the minimum rate results of Fig. 9, is presented in   Fig. 10.  The accuracy is measured by the distance of the randomized solution from the upper bound given by the relaxation, following  the standards of Sec. IV-A and \cite{Sidiropoulos2006,Karipidis2008}. In Fig. 10, the results are also normalized by the value of the upper bound.   According to these results,  the  probability for the $\mathrm{SDR}$  to yield rank-1 solutions is reduced as the error radius increases, for all problems. The accuracy reduction  of the $\mathrm{SDR}$ technique as the channel errors increase  was also reported via simulations in \cite{Zheng2009_TSP}, but  for unicast scenarios. What is more, $\mathcal{F_{RB}}$ yields non rank-1 solutions as the   errors increase, with higher probability than the  $\mathrm{SPC}$ problem. However, 1000 randomizations are sufficient to reduce the inaccuracy of all solutions to less than 7\%, as illustrated in Fig. 10. }
{
  It is therefore concluded that although  the relaxation of the robust formulations does not consistently yield rank-1 solutions, especially for higher values of error radius, the Gaussian randomization can provide solutions with adequate accuracy. Finally, the proposed solutions surpass the performance of existing approaches, in practical per-antenna power constrained settings.    }

 \begin{figure}[h]
 \centering
 \includegraphics[width=1\columnwidth]{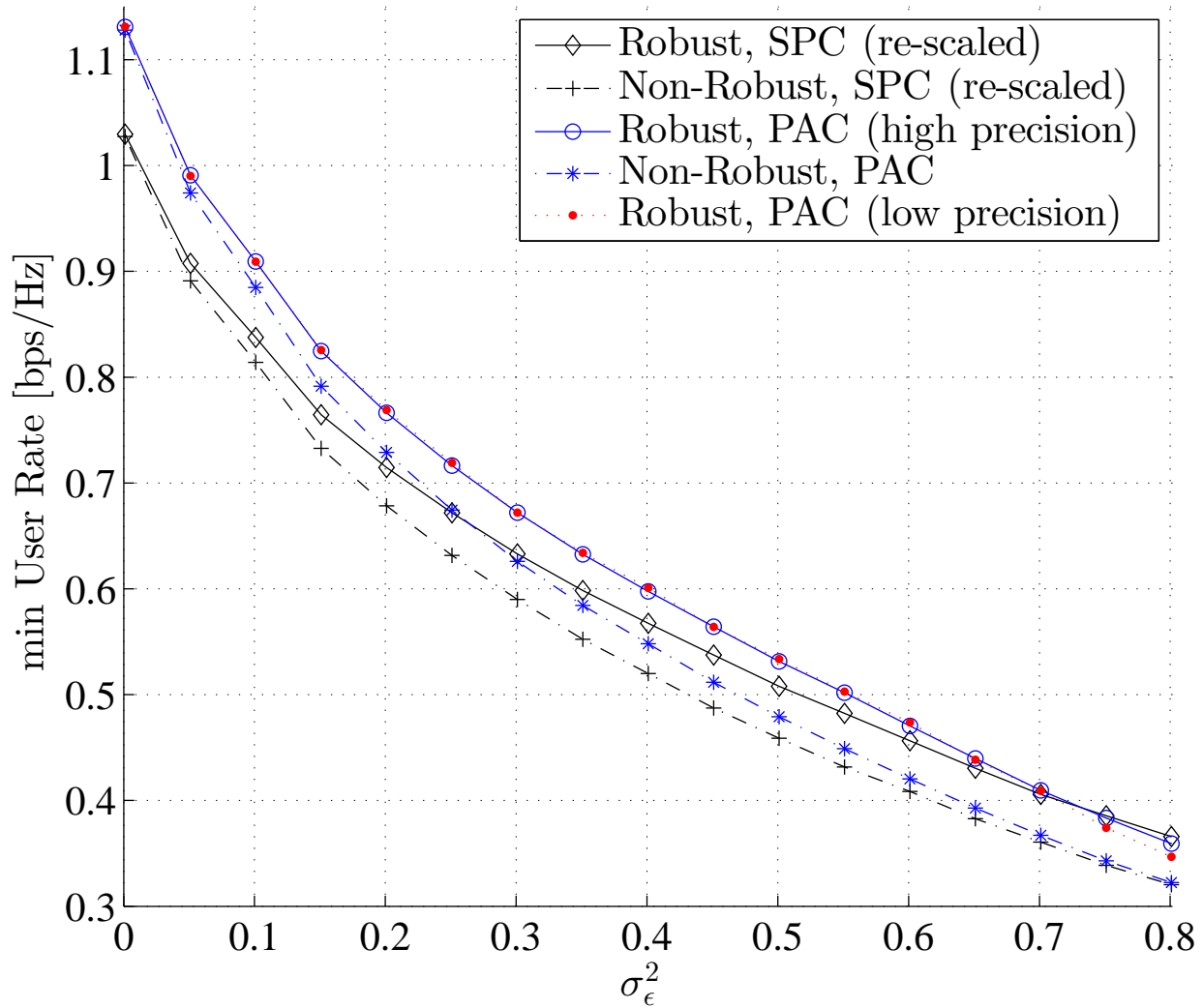}\\
\caption{{Minimum user rate versus increasing $\mathrm{CSI}$ error.} }
\label{fig: robust SR}
\end{figure}
 \begin{figure}[h]
 \centering
 \includegraphics[width=1\columnwidth]{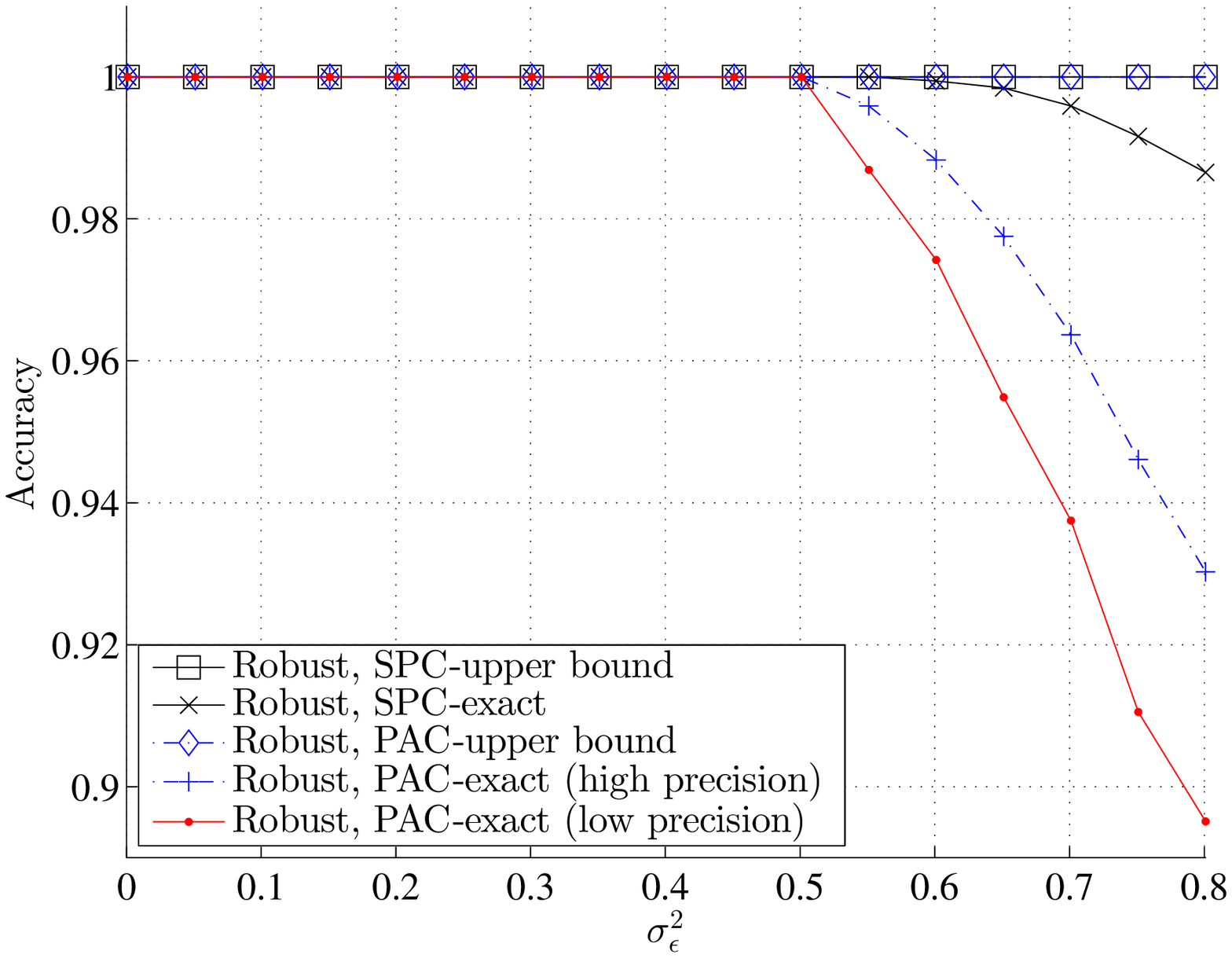}\\
  \caption{{Accuracy of the semidefinite relaxation versus an increasing $\mathrm{CSI}$ error}.
}
\label{fig: robust accuracy}
\end{figure}
%

\section{Conclusions} \label{sec: conclusions}
\textit{}In the present work, optimum linear precoding vectors are derived under per antenna power constraints, when independent sets of common information are transmitted by an antenna array to distinct  co-channel sets of users. The novel weighted max--min fair multigroup multicast problem under $\mathrm{PAC}$s  is formulated. An  approximate solution for this NP-hard  problem is presented based on the well established methods of semidefinite relaxation. 
The performance of the weighted max--min fair multigroup multicast optimization is examined under various system parameters and important insights on the system design are gained. Moreover, an application paradigm of the new system design is described while robust to imperfect $\mathrm{CSI}$ extensions are given.  
Consequently, an important practical constraint towards the implementation of physical layer multigroup multicasting is alleviated.

\bibliographystyle{IEEEtran}
\bibliography{refs/IEEEabrv,refs/conferences,refs/journals,refs/books,refs/references,refs/csi,refs/thesis}

\begin{thebibliography}{10}
\providecommand{\url}[1]{#1}
\csname url@samestyle\endcsname
\providecommand{\newblock}{\relax}
\providecommand{\bibinfo}[2]{#2}
\providecommand{\BIBentrySTDinterwordspacing}{\spaceskip=0pt\relax}
\providecommand{\BIBentryALTinterwordstretchfactor}{4}
\providecommand{\BIBentryALTinterwordspacing}{\spaceskip=\fontdimen2\font plus
\BIBentryALTinterwordstretchfactor\fontdimen3\font minus
  \fontdimen4\font\relax}
\providecommand{\BIBforeignlanguage}[2]{{%
\expandafter\ifx\csname l@#1\endcsname\relax
\typeout{** WARNING: IEEEtran.bst: No hyphenation pattern has been}%
\typeout{** loaded for the language `#1'. Using the pattern for}%
\typeout{** the default language instead.}%
\else
\language=\csname l@#1\endcsname
\fi
#2}}
\providecommand{\BIBdecl}{\relax}
\BIBdecl

\bibitem{Bengtsson2001}
M.~Bengtsson and B.~Ottersten, ``Optimal and suboptimal transmit beamforming,''
  in \emph{Handbook of Antennas in Wireless Communications}.\hskip 1em plus
  0.5em minus 0.4em\relax CRC Press, 2001, pp. 18--1--18--33.

\bibitem{bengtsson1999}
------, ``Optimal downlink beamforming using semidefinite optimization,'' in
  \emph{Proc. of Annual Allert. Conf. on Commun. Control and Computing},
  vol.~37.\hskip 1em plus 0.5em minus 0.4em\relax Citeseer, 1999, pp. 987--996.

\bibitem{Schubert2004}
M.~Schubert and H.~Boche, ``Solution of the multiuser downlink beamforming with
  individual {SINR} constraints,'' \emph{{IEEE} Trans. Veh. Technol.}, vol.~53,
  no.~1, pp. 18--28, 2004.

\bibitem{Yu2007}
W.~Yu and T.~Lan, ``Transmitter optimization for the multi-antenna downlink
  with per-antenna power constraints,'' \emph{{IEEE} Trans. Signal Process.},
  vol.~55, no.~6, pp. 2646--2660, June 2007.

\bibitem{Dartmann2013}
G.~Dartmann, X.~Gong, W.~Afzal, and G.~Ascheid, ``On the duality of the max min
  beamforming problem with per-antenna and per-antenna-array power
  constraints,'' \emph{{IEEE} Trans. Veh. Technol.}, vol.~62, no.~2, pp.
  606--619, Feb 2013.

\bibitem{Christopoulos2013AIAA}
D.~Christopoulos, P.-D. Arapoglou, S.~Chatzinotas, and B.~Ottersten, ``Linear
  precoding in multibeam satcoms: Practical constraints,'' in \emph{Proc. of
  31st {AIAA} {I}nternational {C}ommunications {S}atellite {S}ystems
  {C}onference ({ICSSC})}, Florence, {IT}, Oct. 2013.

\bibitem{Christopoulos2014_ASMS}
D.~Christopoulos, S.~Chatzinotas, and B.~Ottersten, ``Frame based precoding in
  satellite communications: A multicast approach,'' in \emph{Proc. of {IEEE}
  Adv. Satellite Multimedia Systems (ASMS) Conf.}, 2014, submitted.

\bibitem{Sidiropoulos2006}
N.~Sidiropoulos, T.~Davidson, and Z.-Q. Luo, ``Transmit beamforming for
  physical-layer multicasting,'' \emph{{IEEE} Trans. Signal Process.}, vol.~54,
  no.~6, pp. 2239--2251, 2006.

\bibitem{Karipidis2005CAMSAP}
E.~Karipidis, N.~Sidiropoulos, and Z.-Q. Luo, ``Transmit beamforming to
  multiple co-channel multicast groups,'' in \emph{Proc. of 1st Int. Workshop
  on Comput. Adv. in Multi-Sensor Adapt. Process. (CAMSAP)}, 2005, pp.
  109--112.

\bibitem{Karipidis2008}
------, ``Quality of service and max-min fair transmit beamforming to multiple
  co-channel multicast groups,'' \emph{{IEEE} Trans. Signal Process.}, vol.~56,
  no.~3, pp. 1268--1279, 2008.

\bibitem{Gao2005}
Y.~Gao and M.~Schubert, ``Group-oriented beamforming for multi-stream
  multicasting based on quality-of-service requirements,'' in \emph{Proc. of
  1st Int. Workshop on Comput. Adv. in Multi-Sensor Adapt. Process. (CAMSAP)},
  2005, pp. 193--196.

\bibitem{Pesavento2012b}
A.~Schad and M.~Pesavento, ``Max-min fair transmit beamforming for multi-group
  multicasting,'' in \emph{Proc. of Int. ITG Workshop on Smart Ant. (WSA)},
  2012, pp. 115--118.

\bibitem{Silva2009}
Y.~C.~B. Silva and A.~Klein, ``Linear transmit beamforming techniques for the
  multigroup multicast scenario,'' \emph{{IEEE} Trans. Veh. Technol.}, vol.~58,
  no.~8, pp. 4353--4367, 2009.

\bibitem{Xiang2013}
Z.~Xiang, M.~Tao, and X.~Wang, ``Coordinated multicast beamforming in multicell
  networks,'' \emph{{IEEE} Trans. Wireless Commun.}, vol.~12, no.~1, pp.
  12--21, 2013.

\bibitem{Chatzinotas_JWCOM}
S.~Chatzinotas, M.~Imran, and R.~Hoshyar, ``On the multicell processing
  capacity of the cellular {MIMO} uplink channel in correlated {Rayleigh}
  fading environment,'' \emph{{IEEE} Trans. Wireless Commun.}, vol.~8, no.~7,
  pp. 3704--3715, July 2009.

\bibitem{Christopoulos2014_ICC}
D.~Christopoulos, S.~Chatzinotas, and B.~Ottersten, ``Multigroup multicast
  beamforming under per antenna power constraints,'' in \emph{Proc. of {IEEE}
  Int. Commun. Conf.}, 2014, accepted.

\bibitem{Zheng2011a}
G.~{Zheng}, S.~{Chatzinotas}, and B.~{Ottersten}, ``{Generic Optimization of
  Linear Precoding in Multibeam Satellite Systems},'' \emph{{IEEE} Trans.
  Wireless Commun.}, vol.~11, no.~6, pp. 2308 --2320, Jun. 2012.

\bibitem{convex_book}
S.~Boyd and L.~Vandenberghe, \emph{Convex optimization}.\hskip 1em plus 0.5em
  minus 0.4em\relax Cambridge Univ. Press, 2004.

\bibitem{Luo2010}
Z.-Q. Luo, W.-K. Ma, A.-C. So, Y.~Ye, and S.~Zhang, ``Semidefinite relaxation
  of quadratic optimization problems,'' \emph{{IEEE} Signal Processing Mag.},
  vol.~27, no.~3, pp. 20--34, 2010.

\bibitem{ye2011interior}
Y.~Ye, \emph{Interior point algorithms: theory and analysis}.\hskip 1em plus
  0.5em minus 0.4em\relax John Wiley \& Sons, 2011, vol.~44.

\bibitem{Karipidis2006}
E.~Karipidis, N.~Sidiropoulos, and Z.-Q. Luo, ``Convex transmit beamforming for
  downlink multicasting to multiple co-channel groups,'' in \emph{Proc. of IEEE
  Int. Conf. on Acoustics, Speech and Signal Proc. (ICASSP)}, vol.~5, May 2006.

\bibitem{Karipidis2007}
------, ``Far-field multicast beamforming for uniform linear antenna arrays,''
  \emph{{IEEE} Trans. Signal Process.}, vol.~55, no.~10, pp. 4916--4927, Oct
  2007.

\bibitem{Gershman2010}
A.~Gershman, N.~Sidiropoulos, S.~Shahbazpanahi, M.~Bengtsson, and B.~Ottersten,
  ``Convex optimization-based beamforming,'' \emph{{IEEE} Signal Processing
  Mag.}, vol.~27, no.~3, pp. 62--75, 2010.

\bibitem{Shenouda2007}
M.~Shenouda and T.~Davidson, ``Convex conic formulations of robust downlink
  precoder designs with quality of service constraints,'' \emph{{IEEE} J.
  Select. Topics Signal Process.}, vol.~1, no.~4, pp. 714--724, Dec. 2007.

\bibitem{Jindal2004}
N.~Jindal, S.~Vishwanath, and A.~Goldsmith, ``On the duality of {Gaussian}
  multiple-access and broadcast channels,'' \emph{{IEEE} Trans. Inf. Theory},
  vol.~50, no.~5, pp. 768--783, May 2004.

\bibitem{Chen2012}
Z.~Chen, W.~Zhang, and G.~Wei, ``Robust transmit beamforming for multigroup
  multicasting,'' in \emph{IEEE Vehic. Tech. Conf. (VTC Fall)}, Sept 2012, pp.
  1--5.

\bibitem{Zheng2009_TSP}
G.~Zheng, K.-K. Wong, and B.~Ottersten, ``Robust cognitive beamforming with
  bounded channel uncertainties,'' \emph{{IEEE} Trans. Signal Process.},
  vol.~57, no.~12, pp. 4871--4881, Dec. 2009.

\end{thebibliography}
\end{document}